\documentclass[aps,prb,twocolumn,groupedaddress,showpacs,amsmath]{revtex4}

\usepackage{graphicx} \usepackage{latexsym}

\newcommand{\cats}[1]{\ensuremath{\left|#1\right>}}
\newcommand{\bras}[1]{\ensuremath{\left<#1\right|}}
\newcommand{\up}{\ensuremath{\uparrow}}
\newcommand{\down}{\ensuremath{\downarrow}}
\newcommand{\tr}{\ensuremath{\operatorname{tr}}}
\newcommand{\im}{\ensuremath{\mathrm{i}}}
\newcommand{\Ham}{\ensuremath{\mathcal{H}}}
\newcommand{\arsinh}{\ensuremath{\operatorname{arsinh}}}
\newcommand{\e}{\ensuremath{\mathrm{e}}}

\begin{document}

\preprint{}

\title{Thermodynamic properties and thermal correlation lengths \\ of
  a Hubbard model with bond-charge interaction}

\author{Andreas Kemper} \email[]{kemper@thp.uni-koeln.de}
\author{Andreas Schadschneider} \email[]{as@thp.uni-koeln.de}
\affiliation{Institute for Theoretical Physics, University of Cologne,
  50937 K\"oln, Germany} \date{\today}

\begin{abstract}
  We investigate the thermodynamics of a one-dimensional Hubbard model
  with bond-charge interaction $X$ using the transfer matrix
  renormalization group method (TMRG). Numerical results for various
  quantities like spin and charge susceptibilities, particle
  densities, specific heat and thermal correlation lengths are
  presented and discussed. We compare our data also to results for the
  exactly solvable case $X/t=1$ as well as to bosonisation results for
  weak coupling $X/t\ll 1$, which shows excellent agreement. We
  confirm the existence of a Tomonaga-Luttinger and a Luther-Emery
  liquid phase, in agreement with previous studies at zero
  temperature.  Thermal singlet-pair correlation lengths are shown to
  dominate density and spin correlations for finite temperatures in
  certain parameter regimes.
\end{abstract}

\pacs{71.10.Fd, 71.27.+a, 5.10.Cc, 5.30.$-$d, 5.70.$-$a}

\maketitle


\section{Introduction}
The Coulomb interaction between electrons in a solid does not only
lead to diagonal terms like the Hubbard interaction.  It is also
responsible for other non-diagonal terms that are often neglected
since the corresponding coupling constants are assumed to be small.
However, this is not always true. Hirsch (see e.g.\ refs.\ 
\onlinecite{H02,H03} and references therein) has argued that
especially the bond-charge interaction $X$ may be responsible for
interesting physics based on the breaking of particle-hole symmetry.

In refs.~\onlinecite{H89,H89_2,MH90} it has been shown within a
BCS-type analysis that the bond-charge interaction can induce an
effective attraction between particles at densities $n>1$, which can
even lead to the formation of Cooper pairs and superconductivity.
More evidence for this scenario has been obtained in analytical
\cite{BKSZ,JM94,JM94_2,Buzatu,Bulka98,AGAH98,JK99} and numerical
\cite{QSZ95,Q95,AAGHB94,AHBG00} studies of one-dimensional variants
and an effective model for CuO$_2$ planes \cite{AGP93}.  Other
materials where bond-charge interaction might play an important role
are polyacetylene (see ref. \onlinecite{CGL90} and references therein)
and the Bechgaard salts \cite{AG99}.

The Bechgaard salts show a rich phase diagram as temperature and
pressure are varied \cite{BJ99}. At temperatures not too low these
materials can be regarded as quasi one-dimensional. In order to
elucidate the relevance of bond-charge interactions it is therefore
interesting to study thermodynamic properties of one-dimensional
generalized Hubbard models with bond-charge interactions.  This has
been done in ref.~\onlinecite{DM02} for the integrable case $t=X$
where $t$ is the single-particle hopping matrix. Here the Hamiltonian
simplifies considerably since the number of doubly-occupied sites is
conserved. This allows for an exact solution for the complete spectrum
\cite{AA94,S95,dBKS} which has then been used for the calculation of
the thermodynamics in ref.~\onlinecite{DM02}.  However, it should be
noted that the Hamiltonian at $t=X$ has particle-hole symmetry such
that the physics is different from Hirsch's scenario. We therefore
investigate here the thermodynamics away from the point $t=X$. Since
no exact solution is available one has to rely on numerical methods.
Here we employ the transfer matrix DMRG (TMRG) algorithm.

The TMRG method was originally introduced by Nishino \cite{N95} to
study two-dimensional classical models. An adaption to thermodynamics
of quantum systems was proposed by Xiang et.\ al
\cite{BXG96,WX97,WX99}, who used a Trotter-Suzuki decomposition
\cite{T59,S85} to map the 1D quantum model to a 2D classical one.
Thereby, various thermodynamic properties can be studied with high
precision \cite{KRS99,SK02,SK02_2}. Similar ideas have recently been
applied to stochastic models \cite{KSZ01,KGNSZ}.

The paper is organised as follows. In Sec.~\ref{sec:model} we present
the model and its known properties. In Sec.~\ref{sec:TMRG} the TMRG
approach to the thermodynamics of one-dimensional quantum systems is
explained. Especially we emphasize the extensions necessary due to the
fermionic nature of the problem.  Sec.~\ref{sec:results} contains our
main results. We analyze not only thermodynamic quantities like
specific heat and susceptibilities, but also correlation functions,
expressed by thermal correlation lengths.  Sec.~\ref{sec:concl}
contains a summary and conclusions.


\section{The Model} \label{sec:model}
The Hamiltonian of the one-dimensional bond-charge interaction model
reads
\begin{eqnarray}
  \label{eq:hamiltonian}
  \mathcal H =&&-t\sum_{\left<ij\right>\sigma} c^\dagger_{i\sigma}
  c_{j\sigma} + X\sum_{\left<ij\right>\sigma} c^\dagger_{i\sigma}
  c_{j\sigma} (n_{i\bar\sigma}+n_{j\bar\sigma}) \nonumber \\ 
  &&-h\sum_i (n_{i\up}-n_{i\down})-\mu\sum_i (n_{i\up}+n_{i\down})
\end{eqnarray}
where $c_{i\sigma}^\dagger$ and $c_{i\sigma}$, respectively, are
fermionic creation and annihilation operators of an up- or
down-electron ($\sigma=\uparrow,\downarrow$) at site $i$.  In
eq.~(\ref{eq:hamiltonian}) the index $\left<ij\right>$ denotes the
summation over adjacent lattice sites,
$n_{i\sigma}=c_{i\sigma}^\dagger c_{i\sigma}$ the number of particles
with spin $\sigma$ at site $i$ and $\bar\sigma$ the opposite of spin
$\sigma$. Obviously, the amplitude $t>0$ is modified by $X>0$, which
correlates the hopping process to the charge of opposite spins.

The bond-charge model (\ref{eq:hamiltonian}) exhibits full
$\mathrm{SU}(2)$ spin and $\mathrm U(1)$ charge symmetry. Under the
particle-hole transformation
\begin{equation}
  c_{j\sigma} \to (-1)^j c_{j\sigma}^\dagger
\end{equation}
the Hamiltonian transforms (up to constants) into
\begin{equation}
  \label{eq:Htrafo}
  \mathcal H(t,X,\mu,h) \to \mathcal H(t-2X,-X,-\mu,-h) \ ,
\end{equation}
Thus, we can restrict ourselves to the parameter region $0\le X/t\le
1$, which is representative for the whole model. $0\leq X/t\le 0.5$
coincides with $X/t\leq 0$ such that for weak $X$ an repulsive
correlated hopping amplitude transforms into an attractive one for
holes. $0.5\le X/t\le 1$ conversely corresponds to $X/t\ge 1$.

Generically, two universality classes of critical liquid phases are
found for such two-component systems at temperature $T=0$.  In a
Tomonaga-Luttinger liquid (TLL) phase spin and charge excitations are
gapless and all correlation functions decay algebraically. In
contrast, a Luther-Emery liquid (LEL) has a gapped spectrum of spin
excitations and criticality is only observed for charge excitations.

For $X/t\ll1$ the bond-charge model (\ref{eq:hamiltonian}) can be
treated within the framework of bosonisation techniques and
renormalisation methods \cite{JM94}.  It was found, that
(\ref{eq:hamiltonian}) corresponds to an effective Hubbard model with
\begin{equation}
  U_\text{eff}=8X\cos(k_F) \ ,
\label{eq_Ueff}
\end{equation}
and effective hopping amplitude
\begin{equation}
  t_\text{eff}=(1-nX)t
\end{equation}
where $n$ is the filling and $k_F=n \pi/2$.  Thus, two phases appear
for $X\ll 1$ corresponding to the effective Hubbard model: a TLL
regime (repulsive Hubbard interaction) 
below half filling ($n<1$) and a superconducting LEL (attractive
Hubbard interaction) for $n>1$.
It is a crucial question, if the range of applicability of the
bosonisation approach holds (qualitatively) for $X\sim t$. At least we
know, that the integrable case $X/t=1$ contradicts with the
bosonisation results.

For $X/t=1$ the bond-charge model (\ref{eq:hamiltonian}) shows an
additional $\mathrm{SU}(2)$ pseudospin symmetry induced by the
so-called $\eta$-pair operators \cite{Yang89}
\begin{eqnarray}
  \eta^\dagger &=& \sum_i c_{i\downarrow}^\dagger
  c_{i\uparrow}^\dagger, \quad \eta = \sum_i c_{i\uparrow}
  c_{i\downarrow}, \notag \\ \eta^z &=&
  \frac{1}{2}\sum_i(1-c_{i\uparrow}^\dagger c_{i\uparrow} -
  c_{i\downarrow}^\dagger c_{i\downarrow}) \ ,
\end{eqnarray}
and is particle-hole invariant due to eq.~(\ref{eq:Htrafo}).  Apart
from that, the number of double occupied sites
$N_{\uparrow\downarrow}=\sum_i n_{i\uparrow} n_{i\downarrow}$ is also
conserved \cite{S95}.

In that case the model is solvable by mapping the Hamiltonian to free
spinless fermions \cite{AA94,S95}.  One can observe the complete
energy spectrum of $\mathcal H$ and distinguish three different $T=0$
phases.  At intermediate fillings $0.5<n<1.5$ some of the ground
states are of $\eta$-pairing type, which show ODLRO and hence are
superconducting \cite{Yang89,Yang62} (see, however,
ref.~\onlinecite{AAG96}).  For small fillings $n<0.5$ the ground
states are identical to those of the $U=\infty$ Hubbard model.  Due to
particle-hole symmetry of the model, a corresponding phase occurs for
large fillings $n>1.5$.  The model exhibits no spin gap, thus all
phases fall into the TLL universality class.

The knowledge of the complete energy spectrum can be used to calculate
the partition function $Z=\tr \e^{-\beta \mathcal H}$ exactly
\cite{DM02}. This leads to the grand canonical potential
\begin{equation}
  \label{eq:phi}
  -\beta \phi = \ln\big(1+\e^{2\beta\mu}\big) + \frac{1}{\pi}
  \int_0^\pi \ln\big(1+\e^{-\beta(\epsilon_k -
    \mu^*)}\big)\,\mathrm{d}k
\end{equation}
with $\epsilon_k=-2t\cos k$ and the effective chemical potential
\begin{equation}
  \mu^*(\mu,\beta,h) = \mu + \frac{1}{\beta}\ln \frac{2\cosh \beta h}
  {1+\e^{2\beta\mu}} \ .
\end{equation}
The potential $\phi$ can be used to calculate various thermodynamic
properties of the model rigorously \cite{DM02}.

For the non-integrable regime $0<X/t<1$ detailed numerical studies of
ground state properties and correlation functions are available
\cite{JM94,Q95,QSZ95,AAGHB94}.  It was shown that the model exhibits a
spin gap for $0<X/t\lesssim0.75$ and sufficiently large fillings.
Conversely, the spin gap closes for $X/t\gtrsim0.75$ and TLL behaviour
is found for all fillings $n$.

A focus of refs.~\onlinecite{Q95,QSZ95} was set on various two-point
correlation functions, such as density, spin, singlet and triplet pair
correlations. It was found that pair correlations are dominant for the
spin gap regime $0<X\lesssim 0.75$. $X/t\approx 0.5$ was identified as
a particular point where the spin gap and the dominance of
superconducting correlations are maximal.

The present paper expands the $T=0$ studies to \emph{finite
  temperatures} $T>0$. Apart from the investigation of
susceptibilities and the specific heat, we particularly focus on
thermal correlation functions, especially the possibility of 
dominant superconducting pair correlations at $T>0$.  For the
numerical analysis we use the transfer matrix DMRG (TMRG) method,
which has been shown to be highly precise for fermionic models
\cite{SK02,SK02_2} without suffering from minus-sign problems etc.
There it has also been demonstrated, that the asymptotic behaviour of
various correlation functions in terms of \emph{thermal correlation
  lengths} are accessible by the TMRG algorithm.


\section{The TMRG Method} \label{sec:TMRG}
The transfer matrix DMRG (TMRG) method, a variant of White's DMRG
algorithm \cite{W92,W93}, has originally been introduced by Nishino to
study two-dimensional classical systems \cite{N95}. Xiang et.\ al
applied the TMRG to the quantum transfer matrix (QTM) to study the
thermodynamics of one-dimensional quantum systems
\cite{BXG96,WX97,WX99}.  Since then it has been successfully used to
study various spin and fermion systems \cite{KRS99,SK02,SK02_2}.

\subsection{Transfer matrix approach}
The key idea of the TMRG is a Suzuki-Trotter decomposition
\cite{T59,S85} of the partition function
\begin{equation}
  \label{eq:trotter}
  Z = \tr \e^{-\beta\Ham} = \tr \big(\e^{-\epsilon \Ham_\mathrm{o}}
  \e^{-\epsilon\Ham_\mathrm{e}}\big)^M + O(\epsilon^2)
\end{equation}
where $\epsilon=\beta/M$ is fixed and has to be chosen sufficiently small.  The
Hamiltonian $\Ham$ consists of next-neighbour interactions
$\Ham=\sum_i h_{i,i+1}$ and is split up into
$\Ham_\mathrm{o/e}=\sum_{\textrm{o/e}} h_{i,i+1}$, where the sum is
taken over odd and even bonds, respectively.
\begin{figure}
  \centering \includegraphics[width=0.99\columnwidth]{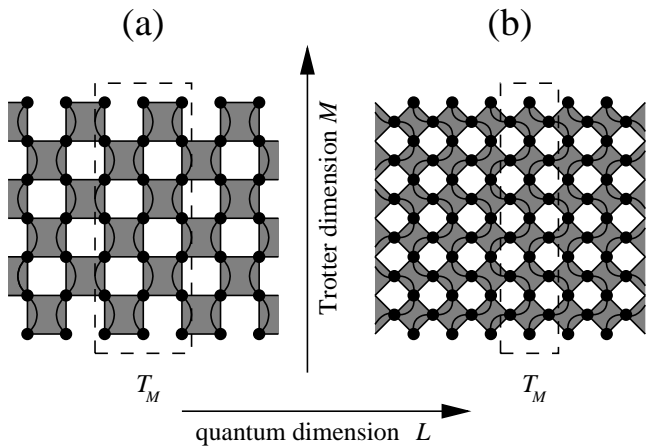}
  \caption{\label{fig:trotter} Suzuki-Trotter decomposition of the
    partition function $Z$, leading to a classical two-dimensional
    lattice. The dots ($\bullet$) represent the lattice sites of
    classical spins, which interact by the shades plaquettes. In (a)
    the traditional mapping is plotted, cf.\ ref.~\onlinecite{S85}.
    An alternative lattice is depicted in (b), which has additional
    auxiliary lattice points, cf.\ ref.~\onlinecite{SK02}. The quantum
    transfer matrix $T_M$ is constructed column-wise.}
\end{figure}
The decomposition leads to a two-dimensional classical model with a
checkerboard structure (Fig.~\ref{fig:trotter}~(a)) and plaquette
interactions $\tau$ given by
\begin{equation}
  \tau_{l_2r_2}^{l_1r_1} \, = \,\,
 \begin{minipage}{1cm}\vspace{6pt}
   \includegraphics[width=1cm]{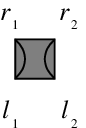}
 \end{minipage}
 \,\, = \, \bras{l_1l_2}\e^{-\beta h_{i,i+1}}\cats{r_1r_2} \ .
\end{equation}
The quantum direction $L$ is thereby supplemented by a virtual Trotter
dimension $M$.  The plaquettes are column-wise combined to the quantum
transfer matrix (QTM) $T_M$, cf.\ Fig.~\ref{fig:trotter}(a).
Alternatively, the decomposition of the partition can be constructed
\cite{SK02} as shown in Fig.~\ref{fig:trotter}(b), which is fully
translational invariant in quantum direction, in contrast to (a).  The
advantage of (b) is discussed later, we first focus on variant (a).

One finds that the thermodynamics in the thermodynamic limit
$L\to\infty$ is determined by the leading eigenvalues $\Lambda_\alpha$
and eigenvectors $\big|\Lambda_\alpha^{L/R}\big>$ $(\alpha=0,1,\dots)$
of $T_M$. Note, that left and right eigenvectors have to be
distinguished because $T_M$ is in general \emph{not} symmetric.  The
free energy explicitly reads
\begin{equation}
  f = -\frac{T}{2} \ln \Lambda_0
\label{eq_free-energy}
\end{equation}
where the leading eigenvalue $\Lambda_0$ is real and unique due to
Frobenius' theorem. Thus, we always observe a gap in $T_M$, which only
may close and lead to a phase transition for $M\to\infty$ ($T\to 0$).
The expectation value of any local operator $O_j$ is given by
\begin{equation}
  \left<O_j\right> =
  \frac{\bras{\Lambda_0^L}T_M(O_j)\cats{\Lambda_0^R}}{\Lambda_0}
\end{equation}
where in $T_M(O_j)$ one plaquette of the QTM $T_M$ has been modified
by the operator $O_j$.

In the TMRG algorithm, the matrices $T_M$ and $T_M(O_j)$ are
sequentially enlarged using a DMRG-like algorithm. An asymmetric
density matrix $\rho=\tr'\cats{\Lambda_0^R}\bras{\Lambda_0^L}$ is used
to target the leading eigenvector. We do not discuss the details,
since the TMRG we use basically follows Ref.~\onlinecite{WX97,WX99}.
As spin and particle number are conserved by $\Ham$, $T_M$ has a block
structure that drastically reduces the computational effort.
Alternatively, we use the number of particles with spin up and down
$N_\sigma=\sum_i n_{i\sigma}$ to build the QTM blocks. To be more
precise, $N_\sigma$ are conserved for a row-transfer matrix.
Contrary, $\Delta N_\sigma=\sum_i (-1)^i n_{i\sigma}$ are the
corresponding good quantum numbers for the (column) quantum transfer
matrix $T_M$.\cite{WX99} Hence, we label the QTM block by the tupel
$(\Delta N_\up, \Delta N_\down)$. The leading eigenvalue $\Lambda_0$
is located in the $(0,0)$ block. This follows directly from the high
temperature limit $T\to\infty$, where $\e^{-\beta h_{i,i+1}}$ is unity
and therefore
\begin{equation}
  \label{eq:tau_0}
  \tau_{l_2r_2}^{l_1r_1} = \delta_{l_2 r_2} \delta_{l_1 r_1} \ .
\end{equation}
Thus, only matrix elements of the $(0,0)$ block of $T_M$ are
non-vanishing. Consequently $\Lambda_0$ is also found in the $(0,0)$
block.  This property persists for finite temperatures $T>0$, because
no level crossing is expected, which would imply a phase transition.

\subsection{Correlation lengths}
\label{sub_corrlength1}
One focus of the present work are correlation functions, such as spin,
density and superconducting pair correlations. The general form of a
two-point correlation function $\big<O_0^\dagger O_j\big>$ for a local
operator $O_j$ at finite temperatures reads
\begin{equation}
  \label{eq:corr_general}
  \big<O_0^\dagger O_j\big> - \big<O_0^\dagger\big>\big< O_j\big> =
  \sum_\alpha M_\alpha \e^{-j/\xi_\alpha} \e^{\im k_\alpha j} \ .
\end{equation}
All correlations decay exponentially because no phase transition can
occur for $T>0$. The correlation lengths (CL) are directly accessible
by the leading eigenvalues
\begin{equation}
  \xi_\alpha^{-1} = \frac{1}{2} \ln \frac{\Lambda_0}{|\Lambda_\alpha|}
  \ .
\label{eq_corr-length}
\end{equation}
The wavevectors $k_\alpha$ are given by the complex argument
\begin{equation}
  k_\alpha = \frac{1}{2} \arg
  \bigg(\frac{\Lambda_0}{\Lambda_\alpha}\bigg) + n \pi \quad (n=0,1) \ 
  .
\end{equation}
As the classical lattice in Fig.~\ref{fig:trotter}(a) is 
translational invariant in quantum direction, but with a doubled unit
cell, $k_\alpha$ is only
determined modulo $\pi$. This problem can be solved by using the QTM
presented in Fig.~\ref{fig:trotter}(b), cf.\ ref.~\onlinecite{SK02}.

The crucial question which eigenvalue $\Lambda_\alpha$ refers to which
correlation function is controlled by the coefficients $M_\alpha$,
that are given by \cite{WX99}
\begin{equation}
  \label{eq:selrule}
  M_\alpha =
  \frac{\bras{\Lambda_0^L}T_M(O_0^\dagger)\cats{\Lambda_\alpha^R}
    \bras{\Lambda_\alpha^L}T_M(O_j)\cats{\Lambda_0^R}}
  {\Lambda_0\Lambda_\alpha} \ .
\end{equation}
Eq.~(\ref{eq:selrule}) represents a selection rule to assign
$\Lambda_\alpha$ to the correct \emph{type} of correlation length.  A
first criterion are the quantum numbers ($\Delta N_\up, \Delta
N_\down$) of $\big|\Lambda_\alpha^{L/R}\big>$: $M_\alpha$ is
non-vanishing only, if $\bras{\Lambda_\alpha^L}$ and
$T_M(O_j)\cats{\Lambda_0^R}$ have the same quantum numbers. To be more
concrete, we discuss now the relevant correlation functions.

Longitudinal spin ($O_j=n_{j\up}-n_{j\down}$) and charge
($O_j=n_{j\up}+n_{j\down}$) correlations are determined by eigenvalues
$\Lambda_\alpha$ of the $(0,0)$ block of $T_M$, since $O_j$ is
diagonal and $T_M(O_j)$ does not change the quantum numbers. They have
to be distinguished by calculating $M_\alpha$ explicitely.

The situation changes for the superconducting singlet and triplet pair
correlations, where $O_j$ in eq.~(\ref{eq:corr_general}) is given by
\begin{eqnarray}
  P^s_j &=& \frac{1}{\sqrt{2}}
  \big(c_{j\up}c_{j+1\down}-c_{j\down}c_{j+1\up}\big), \\ P^{t1}_j &=&
  \frac{1}{\sqrt{2}}
  \big(c_{j\up}c_{j+1\down}+c_{j\down}c_{j+1\up}\big), \\ P^{t2}_j &=&
  c_{j\up}c_{j+1\up}\ , \quad P^{t3}_j = c_{j\down}c_{j+1\down} .
\end{eqnarray}
The pair correlation operators change the quantum numbers.  Therefore,
eigenvalues $\Lambda_\alpha$ of the $(\pm1,\pm1)$ block contribute to
the correlation lengths of $P^s_j$ and $P^{t1}_j$, whereas those of
the $(\pm 2,0)$ and $(0,\pm 2)$ block belong to $P^{t2}_j$ and
$P^{t3}_j$, respectively.  Without magnetic field, the triplet
correlation functions $P^{ti}_j$ are equal. Thus one can unambiguously
distinguish singlet and triplet pair correlation lengths by computing
the eigenvalues $\Lambda_\alpha$ of the $(1,1)$ and $(2,0)$ block
only.

Note, that the present work uses both types of QTMs of
Fig.~\ref{fig:trotter}.  Even if the second variant improves the
calculation of $k_\alpha$, the checkerboard style QTM
(Fig.~\ref{fig:trotter}(a)) is more precise, especially for the
correlation lengths. The reasons for that are technical ones within
the TMRG algorithm.

\subsection{Jordan-Wigner Transformation}
A sophisticated point in the TMRG algorithm are the fermion statistics
which have \emph{not} been considered in the Suzuki-Trotter mapping in
eq.~(\ref{eq:trotter}). Hence we use a Jordan-Wigner transformation
(JWT) \cite{JWT} to map the fermion system onto a spin model. We
define two unitary operators
\begin{eqnarray}
  K_{j\uparrow}&=& \exp\bigg(\mathrm{i}\pi\sum_{i=1}^{j-1}
  \sum_{\sigma=\uparrow,\downarrow} S_{i\sigma}^+S_{i\sigma}^- \bigg),
  \\ K_{j\downarrow}&=& \exp\bigg(\mathrm{i}\pi\sum_{i=1}^{j}
  \sum_{\sigma=\uparrow,\downarrow}S_{i\sigma}^+S_{i\sigma}^-\bigg)
\end{eqnarray}
where $S_{i\sigma}^{-}$ ($S_{i\sigma}^{+}$) are spin-$1/2$ lowering
(raising) matrices of two "spin types" $\sigma=\up,\down$ at
site $i$. These are related to the fermion operators by
\begin{equation}
  \label{eq:JWT}
  c_{j\sigma} = K_{j\sigma} S_{j\sigma}^- \quad \text{and} \quad
  c_{j\sigma}^\dagger = S_{j\sigma}^+ K_{j\sigma} \ .
\end{equation}
These operators fulfill the Fermi statistics.  Inserting the JWT into
the Hamiltonian (\ref{eq:hamiltonian}), the hopping terms $c_i^\dagger
c_j$ modify, whereas the diagonal charge operators $n_{i\sigma}$
transform canonically:
\begin{eqnarray}
  \label{eq:jwt}
  c_{i+1\up}^\dagger c_{i\up} &\to& (-1)^{n_{i\down}} S^+_{i+1\up}
  S^-_{i\up} \nonumber \\ c_{i+1\down}^\dagger c_{i\down} &\to&
  (-1)^{n_{i+1\up}} S^+_{i+1\down} S_{i\down}^- \nonumber \\ 
  n_{i\sigma} &\to& S^+_{i\sigma} S^-_{i\sigma}
\end{eqnarray}
Note that periodic boundary conditions may transform to twisted ones,
but this is irrelevant for our studies where the thermodynamic limit
is performed exactly. We point out that without the JWT the TMRG
algorithm computes wrong results in particular for the singlet
correlation lengths of the $(\pm 1,\pm 1)$ block, although many other
quantities are not affected.


\section{Results} \label{sec:results}
We present numerical TMRG results for the parameter region $0\le
X/t\le 1$, which is representative for all $X/t$, cf.\ 
Sec.~\ref{sec:model}. In all computations we have set $t=1$ and
temperatures are measured in units of $t$.  As we
use a grand canonical description, the chemical potential $\mu$
controls the density $n$ of particles. Note, that the density is
therefore temperature dependent, $n=n(T,\mu)$. If not stated
differently, we use $\epsilon=0.05$ and retain $N=70-100$ states in
the TMRG algorithm. This is already sufficient for highly accurate
results. We typically compute $M\approx 1000$ TMRG iterations, which
correspond to a minimal reachable temperature of $T=0.02$.

\subsection{Thermodynamics} \label{sec:thermodynamics}
In this section we analyze various thermodynamic quantities: the grand
canonical potential $\phi$, local expectation values (e.g.\ local
density $n$), charge and spin susceptibilities $\chi_c$ and $\chi_s$
and the specific heat $c_\mu$.

The grand canonical potential $\phi(T,\mu)$ is directly accessible
through the largest eigenvalue of the QTM, cf.\ Sec.~\ref{sec:TMRG}.
The charge susceptibility
\begin{equation}
  \chi_c(T,\mu) = \frac{\partial n}{\partial \mu} \big|_T
\end{equation}
is obtained by computing the density $n=\langle n_{i\uparrow}+
n_{i\downarrow}\rangle$ of electrons for two different chemical
potentials $\mu-\Delta\mu$ and $\mu+\Delta\mu$.  Then, we use
numerical derivatives
\begin{equation}
  \chi_c(T,\mu) \approx
  \frac{n(T,\mu+\Delta\mu)-n(T,\mu-\Delta\mu)}{2\Delta\mu}
\end{equation}
to approximate $\chi_c$, where typically $\Delta\mu\sim 10^{-2}$ is
chosen.  Similarly, the spin susceptibility
\begin{equation}
  \chi_s\big|_{h=0} = \frac{\partial m}{\partial h} \big|_{\mu,T;h=0}
\end{equation}
is obtained by the numerical derivative of the magnetisation
$m=\langle n_{i\uparrow}-n_{i\downarrow}\rangle$, varying the magnetic
field $h$.

The natural form of the specific heat in a grand canonical formulation
is given by
\begin{equation}
  \label{eq:cmu}
  c_\mu = T \frac{\partial s}{\partial T}\big|_\mu = T
  \frac{\partial^2 \phi}{\partial T^2} \big|_\mu \ ,
\end{equation}
where $s$ denotes the entropy per site.  But usually one finds data
for the specific heat at constant density $n$, which is related to
$c_\mu$ by the thermodynamic relation
\begin{equation}
  \label{eq:cn}
  c_n = T \frac{\partial s}{\partial T}\big|_n = c_\mu - T
  \frac{\partial n}{\partial T}\big|_\mu \frac{\partial \mu}{\partial
    n}\big|_T \ .
\end{equation}
We only determine $c_\mu$ by calculate numerically the second
derivative of the grand canonical potential $\phi$.

\subsubsection{Exactly Solvable Case $X=1$}

First, we show data for the exactly solvable case $X=1$, basically to
check the precision of the TMRG algorithm.
\begin{figure}
  \centering \includegraphics[width=0.95\columnwidth]{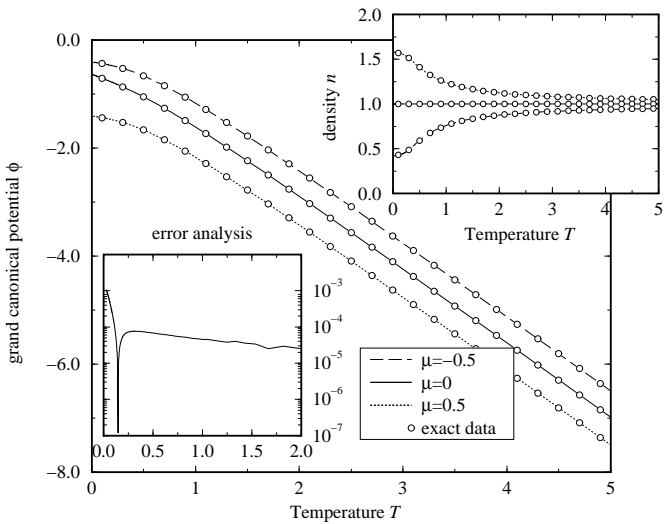}
  \caption{\label{fig:phi} (left) 
    Grand canonical potential $\phi$ as function of temperature for
    $X=1$ and chemical potentials $\mu=-0.5,0,0.5$.  The absolute
    deviation from the exact value in dependence of the temperature is
    plotted in the left inset for $\mu=0$. The corresponding TMRG data
    for the density $n=n(T,\mu)$ is depicted in the right inset. For
    comparison exact data using eq.~(\ref{eq:phi}) are shown by
    circles ($\circ$).}
\end{figure}
In Fig.~\ref{fig:phi} the grand canonical potential $\phi$ is plotted
for the chemical potentials $\mu=-0.5,0,0.5$.  The temperature
dependence of the density $n(T,\mu)$ is plotted in the upper inset.
The particle-hole symmetry of the model is confirmed by the symmetry
of $n(T,\pm 0.5)$ as well as by the potentials $\phi(T,\pm 0.5)$,
which are only shifted by $\Delta\mu=1$.  Our data are compared to
exact results, calculated by eq.~(\ref{eq:phi}), cf.\ the lower inset
of Fig.~\ref{fig:phi}. The absolute error is less than $10^{-4}$ for
temperatures $T>0.1$ and less than $10^{-3}$ for $0.02<T<0.1$. Not
explicitly shown here, the errors of the density $n$ as well as other
expectation values (e.g.\ the magnetisation) are approximately of the
same order.

The origin of a numerical error of at least $10^{-3}$ can be
understood by the approximation done in the Suzuki-Trotter
decomposition in eq.~(\ref{eq:trotter}). The choice of $\epsilon$ (in
our case $\epsilon=0.05$) causes an error of the order $\mathcal O
(\epsilon^2)$ in the partition function (\ref{eq:trotter}).

\begin{figure} 
  \centering \includegraphics[width=0.95\columnwidth]{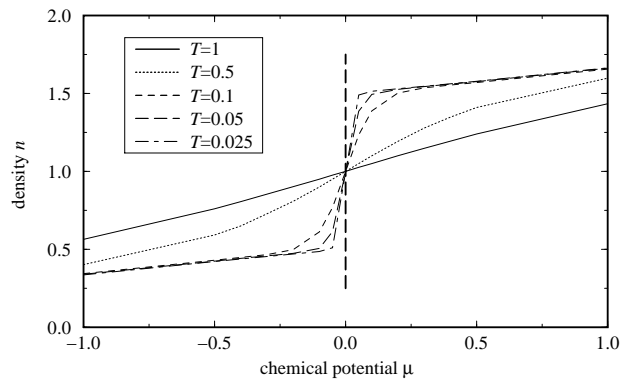}%
  \caption{\label{fig:dens} Density $n(\mu,T)$ as a function
    of temperature and chemical potential for $X=1$. For $T\to 0$ a
     discontinuity at $\mu=0$ is found, where the density  
     jumps from $n=0.5$ to $n=1.5$.}
\end{figure}

It is interesting to investigate also the density $n(T,\mu)$ as a
function of $\mu$ and $T$ (Fig.~\ref{fig:dens}).  For $T\to 0$ one
observes the formation of a plateau for $\mu=0$. This can be explained
by the ground states of the $\eta$-pair phase, which occurs for
$0.5<n<1.5$.  Here, all ground states with different numbers of pairs
$N_{\uparrow\downarrow}$ are degenerate. Thus the chemical potential
does not change by adding a particle, because a new bound pair is
build without changing the energy.
Thus, the thermodynamics of the $\eta$-pair phase can be only realized
for $\mu=0$. Any $\mu\neq 0$ induces $n<0.5$ or $n>1.5$ and falls into
the $U=\infty$ phase, cf.\ Sec.~\ref{sec:model}.

\begin{figure}
  \centering \includegraphics[width=0.95\columnwidth]{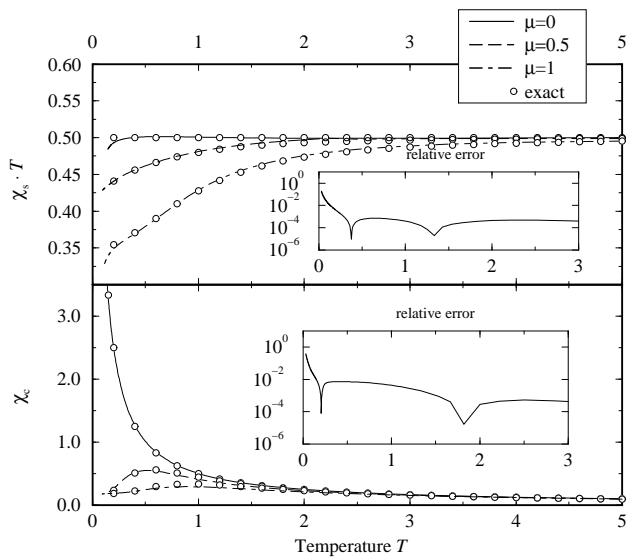}%
  \caption{\label{fig:susz1} Spin and charge susceptibility as a
    function of temperature for $X=1$ and $\mu=0,0.5,1.0$. The upper
    diagram plots $\chi_s \cdot T$ over $T$ in order to demonstrate
    that $\chi_s$ diverges like $\chi_s\sim 1/T$.  The insets depict
    the relative deviation from exact results for $\mu=0$.}
\end{figure}
Fig.~\ref{fig:susz1} depicts the spin and charge susceptibilities for
$X=1$ and $\mu=0,0.5,1.0$.  The comparison with exact data results in
a relative error less than $10^{-2}$ for temperatures $T\gtrsim 0.1$.
Not surprisingly, it is larger than that for $\phi$, because numerical
errors are propagating through the numerical derivatives.

From conformal field theory (CFT) the low temperature limit $T\to 0$
of $\chi_s$ and $\chi_c$ are respectively given by
\begin{equation}
  \label{eq:limitsus}
  \chi_s(T\to0) = \frac{2}{\pi v_s} \quad \text{and} \quad \chi_c(T\to
  0) = \frac{2K_c}{\pi v_c}\ ,
\end{equation}
where $v_s$ and $v_c$ denote the spin and charge velocities and $K_c$
the LL parameter. This can be used to interpret the results of
Fig.~\ref{fig:susz1}. The ground states of $X=1$ for $n<0.5$ (and
after particle-hole transformation for $n>1.5$) corresponds to the
$U=\infty$ Hubbard model, cf. Sec.~\ref{sec:model}.  For the
$U=\infty$ Hubbard model it is known, that
\begin{equation} \label{eq:uinfty}
  v_s=0 \quad \text{and} \quad v_c=2t|\sin(\pi n)|\ .
\end{equation}
The spin excitations exhibit no dispersion due to
complete degeneracy \cite{DM02}. In Fig.~\ref{fig:susz1} we have
plotted $\chi_s\cdot T$ to demonstrate, that the spin susceptibility
$\chi_s$ is diverging for all $\mu$ and fillings $n$ in the limit
$T\to 0$ since $\lim_{T\to 0}T\chi_s$ is finite.  The charge
susceptibility $\chi_c$ is divergent for $\mu=0$.

\subsubsection{Non-Integrable Case $0<X<1$}

We now consider the non-integrable case $0<X<1$.  From analytical
\cite{JM94,JM94_2} and numerical \cite{Q95,QSZ95,AAGHB94} approaches
it is known, that a crossover from a TLL phase to a superconducting
LEL is observed. For $X\ll 1$ this is evident, because the bond-charge
model is an effective Hubbard model (cf.\ Sec.~\ref{sec:model}), where
the crossover is expected to appear at $U_\text{eff}=0$, corresponding
to half-filling $n=1$ (see (\ref{eq_Ueff})).

In our thermodynamic study we can identify the existence of a spin gap
if the susceptibility $\chi_s$ vanishes for $T\to 0$.
\begin{figure}
  \centering \includegraphics[width=0.95\columnwidth]{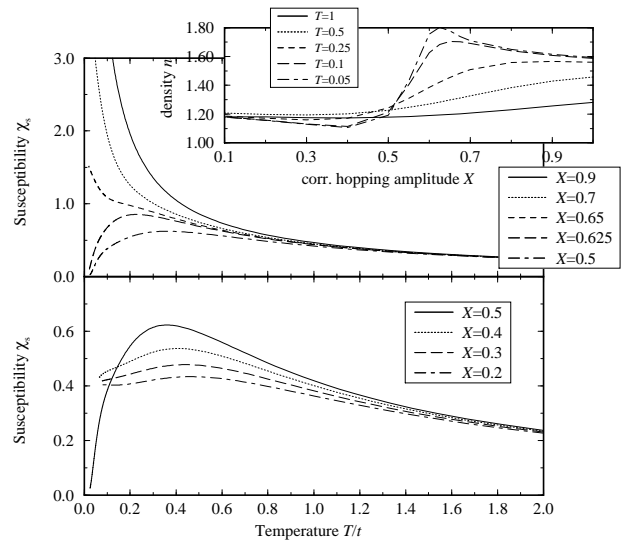}%
  \caption{\label{fig:gap} Spin susceptibility as a function of
    temperature for various amplitudes $X/t=0.9 \dots 0.1$ and fixed
    chemical potential $\mu=0.6$.  The upper figure shows $\chi_s$ for
    $X=0.9\dots0.5$ the lower one for $X=0.1\dots 0.5$, respectively.
    The inset depicts the density $n(X,\mu=0.6,T)$ for various
    parameters $X$.}
\end{figure}
In Fig.~\ref{fig:gap} $\chi_s$ is depicted for fixed chemical
potential $\mu=0.6$ and various hopping amplitudes $0<X<1$. For
$X>0.65$ we can clearly verify TLL behaviour, because the
susceptibility does not vanish for $T\to 0$.  At $X\approx 0.65$ we
observe a crossover to a LEL phase with $\chi_s(T\to 0)=0$. Decreasing
$X$ further, the gap seems to close again for $X\lesssim0.3$.
  
\begin{figure}
  \begin{center}
    \centering \includegraphics[width=0.95\columnwidth]{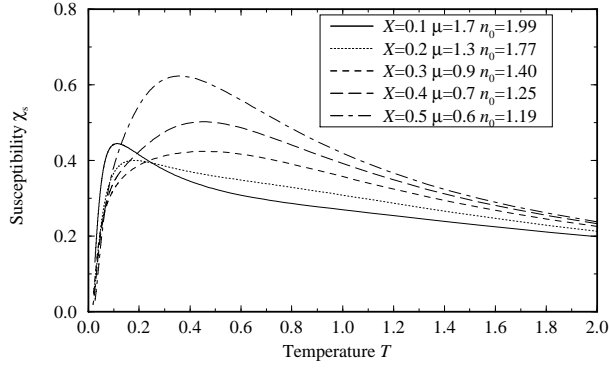}%
    \caption{Spin susceptibility $\chi_s$ as a function 
      of temperature for the parameter region $X=0.5\dots 0.1$ and
      sufficiently large fillings.  The respective densities
      $n_0=n(\mu,T\to 0)$ are given in the legend.}
    \label{fig:gap2}
  \end{center}
\end{figure}

Additional the inset of Fig.~\ref{fig:gap} shows the density
$n(T,\mu=0.6)$ for the respective parameters $X$.  Note, that
$\mu=0.6$ induces a regime of more than half filling $n>1$ for each
$X$. Therefore the spin gap should persist especially for $X\ll 1$,
which we can not show by our TMRG results. But from the attractive
Hubbard model it is known, that the spin gap is exponentially small
for $U\ll 1$.  Since from (\ref{eq_Ueff}) we have $U_\text{eff}\ll 1$
for $X\ll 1$, the gap in the bond-charge model becomes extremely
small, especially at half filling. The decay of $\chi_s$ at $X\ll 1$
is therefore expected at such low temperatures that are not accessible
by TMRG.  To verify that all $X\le 0.5$ exhibit a spin gap we
increased the filling in Fig.~\ref{fig:gap2}.
Here, the effective Coulomb potential $U_\text{eff}$ is larger and the
spin gap appears in our data.

In order to give a more detailed overview of the thermodynamics of the
bond-charge model for $0<X<1$, we focus on three particular points
$X=0.1,0.5,0.9$.

\begin{figure}
  \centering \includegraphics[width=0.95\columnwidth]{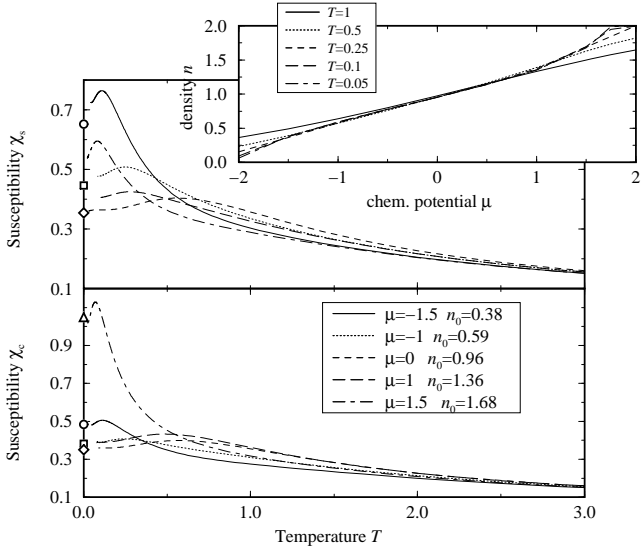}%
  \caption{\label{fig:sus_01} Spin and charge susceptibilities 
    $\chi_s$ and $\chi_c$ as a function of temperature for $X/t=0.1$
    and various chemical potentials $\mu$. The inset plots the density
    $n(\mu,T)$.  The corresponding $T=0$ values for the Hubbard model
    with effective $U_\text{eff}$ and $t_\text{eff}$ are shown by
    symbols, $(\circ):\mu=-1.5$, $(\Box):\mu=-1$, $(\Diamond):\mu=0$
    and $(\triangle):\mu=1.5$.}
\end{figure}

Fig.~\ref{fig:sus_01} plots the charge and spin susceptibility at
$X=0.1$ for various chemical potentials $\mu$. Additionally, in the
inset, the density $n(\mu,T)$ is depicted.
As $X$ is small compared to the bandwidth $t$, the model
(\ref{eq:hamiltonian}) should coincide an the effective Hubbard model.
We check the asymptotics of the susceptibilities $\chi_s$ and $\chi_c$
for $T\to 0$ using eq.~(\ref{eq:limitsus}). These can be calculated
exactly for the integrable Hubbard model \cite{Shiba,Schulz}.  For a
weak (repulsive) Coulomb potential $U$, one finds
\begin{eqnarray}
  \label{eq:hubbardcrit1}
  v_{c/s} &=& v_F \left(1\pm \frac{U}{4\pi t\sin k_F}\right) =v_F \pm
  \frac{U}{2\pi} ,\\ K_c &=& 1 - \frac{U}{4\pi t \sin(k_F)}
\label{eq:hubbardcrit2}
\end{eqnarray}
with $v_F=2t\sin(k_F)$ and $k_F=\frac{n\pi}{2}$.  Using
eqs.~(\ref{eq:hubbardcrit1}), (\ref{eq:hubbardcrit2}) and
(\ref{eq:limitsus}) we have calculated $\chi_s$ and $\chi_c$ for $T\to
0$ (see Fig.~\ref{fig:sus_01}).  The density $n_0=n(\mu,T\to 0)$ which
was used to determine $k_F$, $U_\text{eff}$ and $t_\text{eff}$ is
denoted in the legend of Fig.~\ref{fig:sus_01}. We observe perfect
agreement with our data, which again supports the correspondence of
(effective) Hubbard and bond-charge model.

Even more evidence of Hubbard like thermodynamics is given by the
shape of the specific heat $c_\mu$.
\begin{figure}
  \centering \includegraphics[width=0.95\columnwidth]{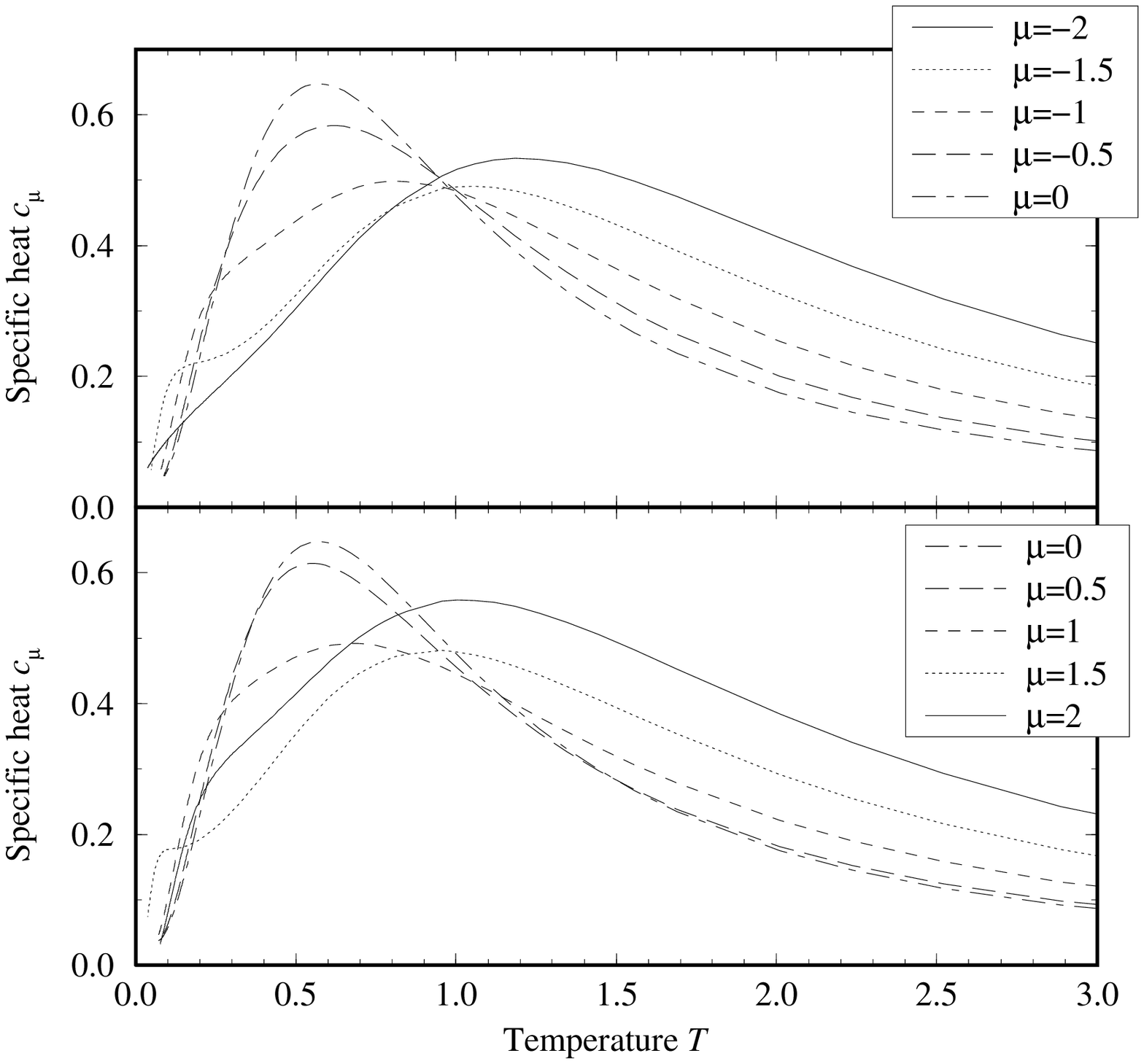}%
  \caption{\label{fig:Cmu_01} Specific heat $c_\mu$ as a function of
    temperature for $X=0.1$ and chemical potentials $\mu\le 0$ (upper
    figure) and $\mu\ge 0$ (lower figure).}
\end{figure}
As shown for $c_\mu$ in Fig.~\ref{fig:Cmu_01}, two characteristic
features can be observed for large and small fillings, a shoulder at
low temperatures and a peak at a slightly larger temperature. As in
the repulsive Hubbard model these features can be related to spin and
charge excitations, respectively. They merge close to half filling
where we have effectively a free fermion system for $X\ll 1$.
Moreover the regime less than half filling $n<1$ closely corresponds
to $n>1$. Again, this can be explained by the exponentially small gap,
which does not affect the physics at those temperature scales we can
observe by TMRG.

\begin{figure}
  \centering \includegraphics[width=0.95\columnwidth]{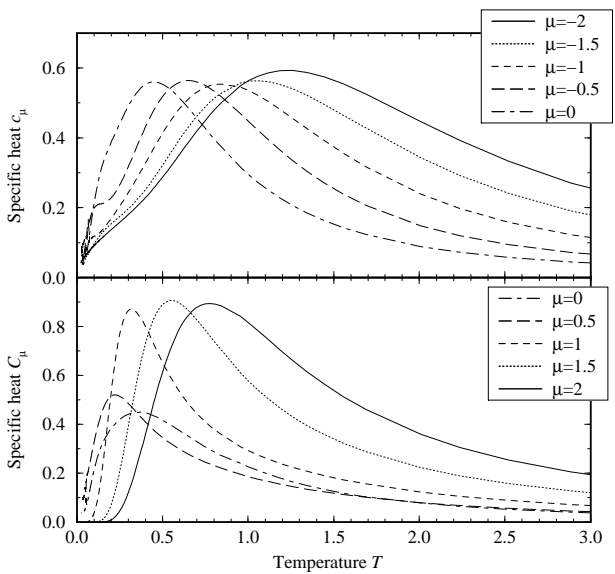}%
  \caption{\label{fig:Cmu_05} Specific heat $c_\mu$ as a function of
    temperature for $X=0.5$ and various chemical potentials $\mu<0$
    (upper figure) and $\mu>0$ (lower figure).}
\end{figure}

The effect of the spin gap manifests itself at larger interactions
$X$.
\begin{figure}
  \centering \includegraphics[width=0.95\columnwidth]{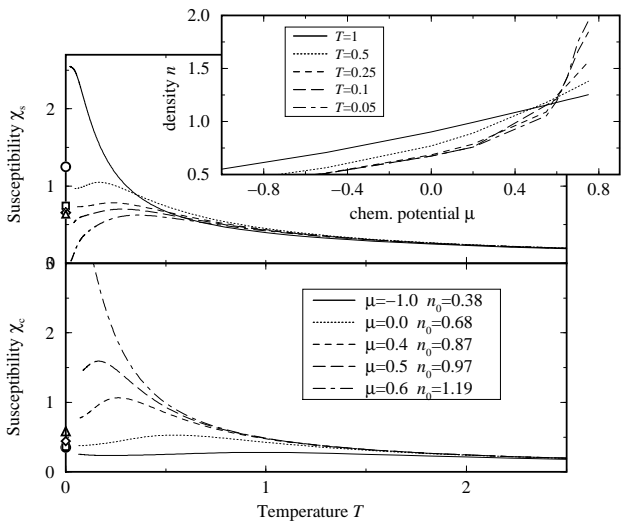}%
  \caption{\label{fig:sus_05} Spin and charge susceptibilities $\chi_s$
    and $\chi_c$ as a function of temperature are depicted for $X=0.5$
    and various chemical potentials $\mu$.  The inset plots the
    density $n(\mu,T)$.  The corresponding $T=0$ values for $\chi_s$
    and $\chi_c$ calculated from the Hubbard model with effective
    $U_\text{eff}$ and $t_\text{eff}$ are shown by symbols,
    $(\circ):\mu=-1$, $(\Box):\mu=0$, $(\Diamond):\mu=0.4$ and
    $(\triangle):\mu=0.5$.}
\end{figure}
Fig.~\ref{fig:sus_05} depicts charge and spin susceptibilites for
$X=0.5$.  In contrast to $X=0.1$, the spin susceptibility $\chi_s$
clearly affirms a gap for $n>1$ since $\chi_s(T\to 0)\to 0$.
Therefore the phase transition occurs at half filling, which is
predicted by the bosonisation results \cite{JM94}. A significantly
larger spin gap than at $X=0.1$ is also verified by the specific heat
shown in Fig.~\ref{fig:Cmu_05}.
The linear behaviour of $c_\mu$ for $\mu\gtrsim 1$ occurs only at very
low temperatures and can not be observed on the temperature scale
shown. Here the exponential corrections coming from the finite spin
gap dominate.

Even though the numerical data for the thermodynamics suggest Hubbard
like behaviour on a qualitative level, a more detailed quantitative
comparison fails for $X=0.5$. Since $U_\text{eff}$ is not small here,
we used the Bethe Ansatz (instead of eq.~(\ref{eq:hubbardcrit1}) and
(\ref{eq:hubbardcrit2})) to calculate the $T\to 0$ values of $\chi_s$
and $\chi_c$ for the (effective) Hubbard model, which are also
depicted in Fig.~\ref{fig:sus_05}. Obviously, these do not fit to our
TMRG data.

\begin{figure}
  \centering \includegraphics[width=0.95\columnwidth]{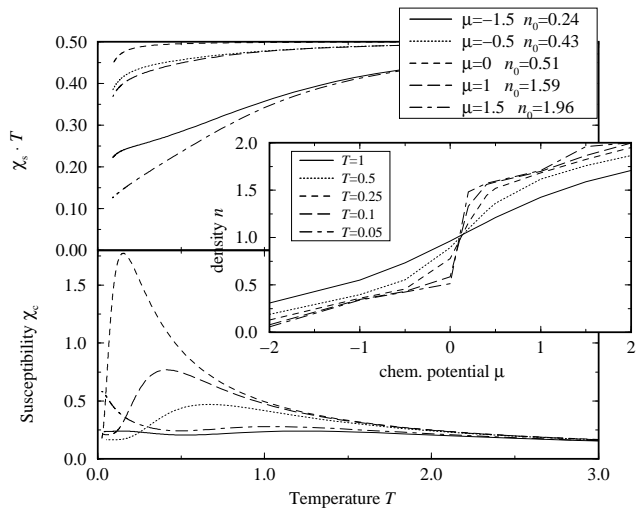}%
  \caption{\label{fig:sus_09} Spin and charge susceptibilities
    $\chi_s$ and $\chi_c$ are
    depicted for $X=0.9$ and various chemical potentials $\mu$. 
    The inset shows the density $n(\mu,T)$.}
\end{figure}

For large $X\approx 1$ the spin gapped phase disappears, which also
contradicts the results of bosonisation. 
As an example we show TMRG data for $X=0.9$. The susceptibilities
and the density, which are shown in Fig.~\ref{fig:sus_09},
qualitatively coincide with the $X=1$ case. 
At $\mu\approx 0$ we observe a jump in the density, indicating
a phase comparable to the $\eta$-pair phase. 
The spin susceptibility $\chi_s(T\to 0)$ is
diverging for all fillings $n$, which yields $v_s\to 0$. 
\begin{figure}
  \centering \includegraphics[width=0.95\columnwidth]{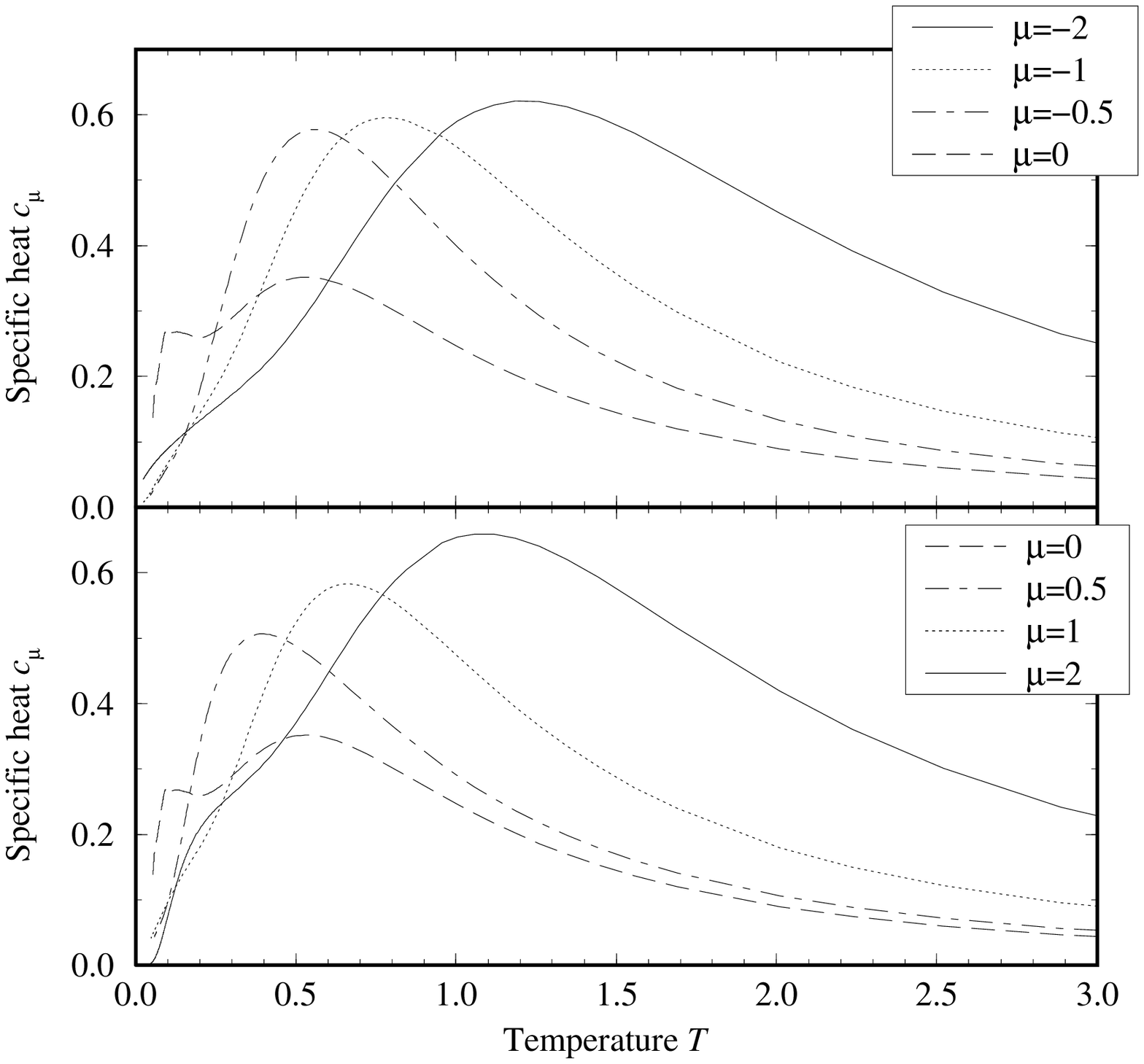}%
  \caption{\label{fig:Cmu_09} Specific heat $c_\mu$ as a function of
    temperature for $X=0.9$ and various
    chemical potentials $\mu\le 0$ (upper figure) and $\mu\ge 0$
    (lower figure).}
\end{figure}
Fig.~\ref{fig:Cmu_09} additionally plots the specific heat $c_\mu$ for
various chemical potentials $\mu$. In fact the model is observed to be
nearly particle-hole symmetric ($\mu\to -\mu$). 
Note, that for $\mu=0$ the specific
heat exhibits an additional low energy peak. Spin excitations can 
not contribute to that peak due to complete degeneracy. 
For $X=1$ such peak has also been observed in
ref.~\onlinecite{DM02}. It is associated with the melting
of pairs. Remarkably, the physics at $X=0.9$ is very similar to the 
highly symmetric $X=1$ case. Thus, $X=1$ is not a
singular point, but the characteristics persist for a certain
neighbourhood around $X=1$. 

\subsubsection{Universal Crossing Points}

In the thermodynamic data shown above we find various 
phenomena of (nearly) universal
crossing points. As a first example, the specific heat in 
Fig.~\ref{fig:Cmu_01}
exhibits even two crossing points. But also in the density in
Fig.~\ref{fig:sus_01} or 
Fig.~\ref{fig:sus_05} a crossing point is apparent at $n\approx 1$.

In the case of the Hubbard model, it is well known \cite{JKS98} that 
such crossing points appear. In refs.~\onlinecite{V97,CMV99} general 
thermodynamic arguments for the existence of a sharp crossing point 
in the specific heat are given. 

Sharp universal crossing points $(\mu_c,n_c)$
in the density $n=n(T,\mu)$ can be generally explained 
in a similar way. The integral
\begin{eqnarray}
  \int_{-\infty}^\infty \frac{\partial n}{\partial T}\big|_\mu \mathrm
  d \mu
  &=& \int_{-\infty}^\infty \frac{\partial s}{\partial \mu}\big|_T
  \mathrm d \mu \notag \\ 
  &=& s(\infty,T)-s(-\infty,T) = 0
\end{eqnarray}
vanishes: for $\mu\to\pm\infty$ the band is either completely empty or
filled, thus $s\to 0$. Then  $\frac{\partial n}{\partial T}\big|_\mu$
has to change its sign. Due to continuity arguments we conclude the
existence of $\mu_c$, such that
\begin{equation}
  \frac{\partial n}{\partial T}\big|_{\mu_c} = 0 \implies
  n(T,\mu_c)=\mathrm{const} . \
\end{equation}

In the case of the bond-charge model at weak coupling $X\ll 1$, 
the crossing point of the density
appears at half filling $n_c=1$, where the model effectively
behaves as free fermions ($U_\text{eff}=0$). 

\subsection{Correlation Lengths}
\label{sec:correlations}

A great advantage of the TMRG algorithm is that not only quantities
related to the free energy (\ref{eq_free-energy}) are accessible, but
also, through the next-leading eigenvalues of the transfer matrix,
thermal correlation lengths (\ref{eq_corr-length}). As explained
in Sec.~\ref{sub_corrlength1}, the corresponding correlation functions
are identified by their quantum numbers $(\Delta N_\uparrow,
\Delta N_\downarrow)$. It is even possible to distinguish the 
contributions of different wavevectors $k_\alpha =\alpha k_F$
$(\alpha =0,2,4,\ldots)$. This will be done in the following to
determine the dominant correlations in the different parameter regimes.
First we discuss the low-temperature behaviour, which can be compared
to predictions of conformal field theory.

\subsubsection{Low-temperature behaviour}
\label{xi-smallT}

The correlation functions of a TLL at temperature $T=0$ show universal 
behaviour. According to conformal field theory 
(CFT), density-density (d-d), spin-spin (s-s), 
singlet pair (sp) and triplet pair (tp) correlation functions read 
\cite{OgSh,JM94}
\begin{eqnarray}
  \left<n_r n_0\right> &\sim&
  A_0 r^{-2} + A_1r^{-(1+K_c)}\cos(2k_F r) \nonumber \\
  && + A_2 r^{-4K_c} \cos(4k_Fr)\ ,\\
  \left<S^z_r S^z_0\right> &\sim&
  B_0 r^{-2} + B_1r^{-(1+K_c)}\cos(2k_F r)\ , \qquad \\
  \left<P^{s\dagger}_r P^s_0\right> &\sim&C_0r^{-(1+\frac{1}{K_c})}
  \nonumber \\
  && + C_1r^{-(K_c+\frac{1}{K_c})}\cos(2k_F r) \ , \\
  \left<P^{t\dagger}_r P^t_0\right> &\sim&D_0r^{-(1+\frac{1}{K_c})}
  \nonumber \\
  && + D_1r^{-\left(K_c+\frac{1}{K_c}+2\right)} \cos(2k_F r) 
\end{eqnarray}
(up to logarithmic corrections).
Thus, all critical exponents are controlled by a dimensionless
parameter $K_c$. 

In contrast, in a LEL the spin excitations are gapped.
Thus s-s and tp correlations decay exponentially
and do not show universal behaviour. According to CFT, the asymptotic
behaviour of the d-d and sp correlations is given by \cite{OgSh,JM94}
\begin{eqnarray}
 \left<n_r n_0\right> &\sim&
  A_0 r^{-2} + A_1r^{-K_c}\cos(2k_F r) \nonumber \\
  && + A_2 r^{-4K_c} \cos(4k_Fr)\ , \\
  \left<P^{s\dagger}_r P^s_0\right> &\sim& C_0r^{-\frac{1}{K_c}} \nonumber \\
  && +C_1r^{-(K_c+\frac{1}{K_c})}\cos(2k_F r) \ .
\end{eqnarray}

In the present work we are discussing correlation functions at
\emph{finite} temperatures $T>0$. Here, the model is always
non-critical and all correlation functions decay
exponentially. The asymptotics are described in terms of thermal
correlations lengths (CL), cf.\ eq.~(\ref{eq:corr_general}), which are
temperature dependent. 

The asymptotic behaviour of the CL for small temperatures $T\to 0$ can 
still be obtained by CFT \cite{SK02_2}. For a TLL all CLs diverge 
\begin{equation}
  \label{eq:xidiverg}
  \xi(T\to 0) = \frac{1}{2\pi T\left(\frac{x_c}{v_c} + \frac{x_s}{v_s}\right)}
    =: \frac{\gamma}{T} \ .
\end{equation}
The scaling dimensions $x_s$ and $x_c$ can be calculated explicitly.
One finds, that the non-oscillating d-d (s-s) CL is given by
\begin{equation}
  \label{eq:dens_corr_0}
  \gamma_\text{dd}^{(0)} = \frac{v_c}{2\pi},\qquad 
 \gamma_\text{ss}^{(0)}=\frac{v_s}{2\pi}\ .
\end{equation}
The $2k_F$ parts of both read
\begin{equation}
  \label{eq:dens_corr_2}
  \gamma_\text{dd/ss}^{(2k_F)} = \frac{v_c}{\pi\left(\frac{v_c}{v_s}+K_c
\right)}
\end{equation}
whereas the $4k_F$ part of the d-d CL yields
\begin{equation}
  \label{eq:dens_corr_4}
  \gamma_\text{dd}^{(4k_F)} = \frac{v_c}{4\pi K_c}\ .
\end{equation}
The non-oscillating parts of sp and tp CL are given by
\begin{equation} \label{eq:sp_corr_0}
  \gamma_\text{sp/cp}^{(0)} = \frac{v_c}{\pi\left(\frac{1}{K_c}
  +\frac{v_c}{v_s}\right)} \ , 
\end{equation}
the $2k_F$ part of the sp CL by
\begin{equation}
  \label{eq:sp_corr_2}
  \gamma_\text{sp}^{(2k_F)} = \frac{v_c}{\pi\left(\frac{1}{K_c}+K_c\right)} 
\end{equation}
and the $2k_F$ part of the tp CL by
\begin{equation}
   \gamma_\text{tp}^{(2k_F)} = \frac{v_c}{\pi\left(\frac{1}{K_c}+K_c
       + 2\frac{v_c}{v_s}\right)} \ .
\end{equation}

In the LEL the tp and s-s CLs do not diverge for $T\to 0$ due to the
spin gap. But the asymptotic behaviour of the d-d and sp CLs is still 
predictable by CFT
and corresponds to eq.~(\ref{eq:dens_corr_0})-(\ref{eq:sp_corr_2})
in the limit $v_s\to \infty$. Thus, we have
\begin{equation} \label{eq:LEL_1}
  \gamma_\text{dd}^{(0)} = \frac{v_c}{2\pi}, \quad 
  \gamma_\text{dd}^{(2k_F)} =\frac{v_c}{\pi K_c} \quad
  \text{and} \quad 
  \gamma_\text{dd}^{(4k_F)} = \frac{v_c}{4\pi K_c} \ 
\end{equation}
for the non-oscillating, $2k_F$ and $4k_F$ d-d CLs and 
\begin{equation} \label{eq:LEL_2}
  \gamma_\text{sp}^{(0)} = \frac{v_c K_c}{\pi} \quad \text{and}
  \quad \gamma_\text{sp}^{(2k_F)} = \frac{v_c}{\pi\left(\frac{1}{K_c}
        +K_c\right)} 
\end{equation}
for the non-oscillating and $2k_F$ sp CL, respectively.
Note, that the low temperature asymptotics of the correlation lengths 
depend not only on the universal exponent $K_c$, but additionally on
the charge and spin velocities $v_c$ and $v_s$.

\subsubsection{Comparison of correlation lengths}

\begin{figure}
  \centering \includegraphics[width=0.95\columnwidth]{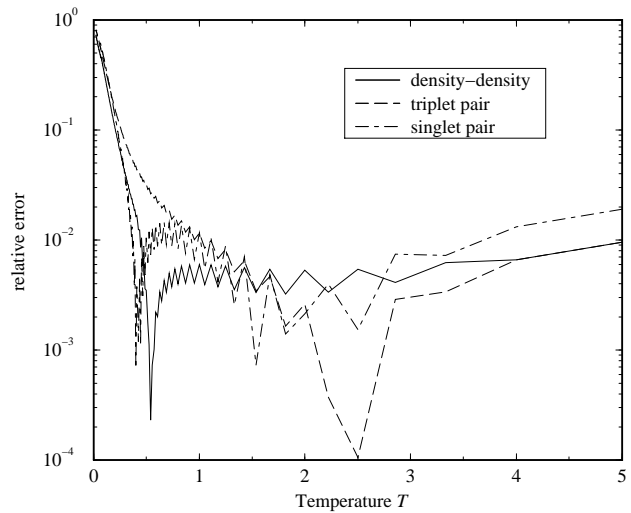}%
  \caption{\label{fig:ff} Precision check of various correlation lengths 
    computed by the TMRG algorithm for free fermions. 
    The relative error is plotted 
    as a function of temperature $T$, showing that the CLs keep
    reliable for $T>0.2$.}
\end{figure}
It was shown in Sec.~\ref{sec:TMRG} that the TMRG algorithm provides a
very effective facility to compute thermal correlation lengths
numerically.
First, we check the precision of our TMRG data. For that purpose we choose
$X=0$, which describes a system of free fermions with spin. 
Here, all correlation lengths can be computed exactly. Due to Wick's theorem,
spin, charge and pair CLs are identical given by \cite{SK02_2}
\begin{equation}
  \xi^{-1} = 2\arsinh(\pi T/2) \ .
\end{equation}

Fig.~\ref{fig:ff} compares the CLs computed by the TMRG program 
to exact data. Down to a temperature $T\approx 0.2$ the relative error is
shown to be less then $10\%$. As already mentioned in
Sec.~\ref{sec:TMRG} it is important to perform a JWT before applying
the TMRG to the bond-charge model. Otherwise, particularly the singlet pair
correlations are \emph{not} correctly reproduced by the algorithm.

We now discuss in detail s-s, d-d, sp and tp correlations 
in the bond-charge model. We focus on the parameter point $X=0.5$, where
the spin gap is comparatively large for $n\gtrsim 1$. For $T=0$, the
system is then a LEL with dominating superconducting sp correlations 
due to $K_c>1$. \cite{QSZ95}
From eqs.~(\ref{eq:LEL_1}) and (\ref{eq:LEL_2}) it is expected, 
that sp correlations should dominate even for finite low temperatures $T>0$. 
At zero temperature, close to half-filling a transition to a TLL
takes place which exists for all densities $n\lesssim 1$.
\begin{figure}
  \centering \includegraphics[width=0.95\columnwidth]{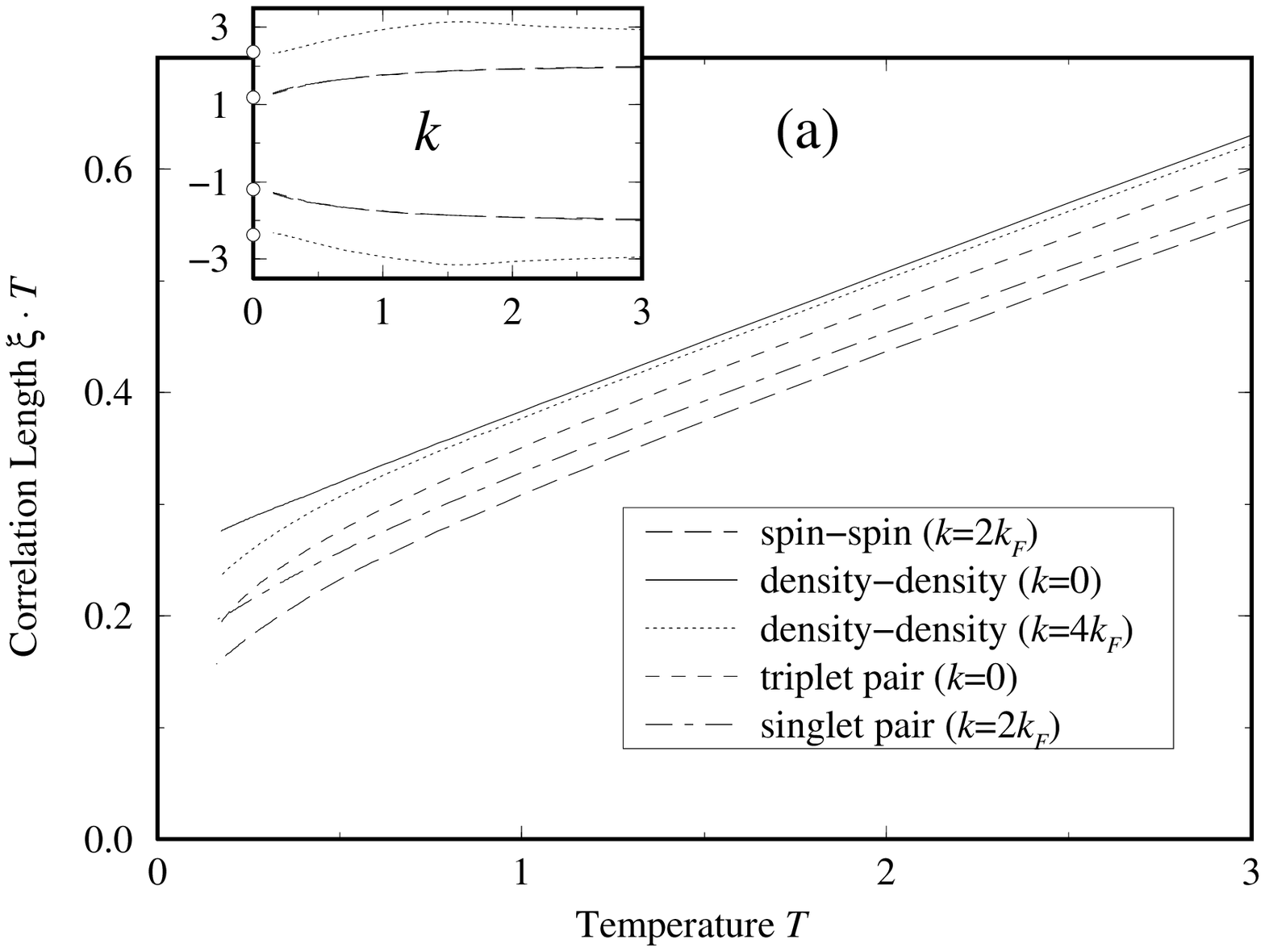}
  \includegraphics[width=0.95\columnwidth]{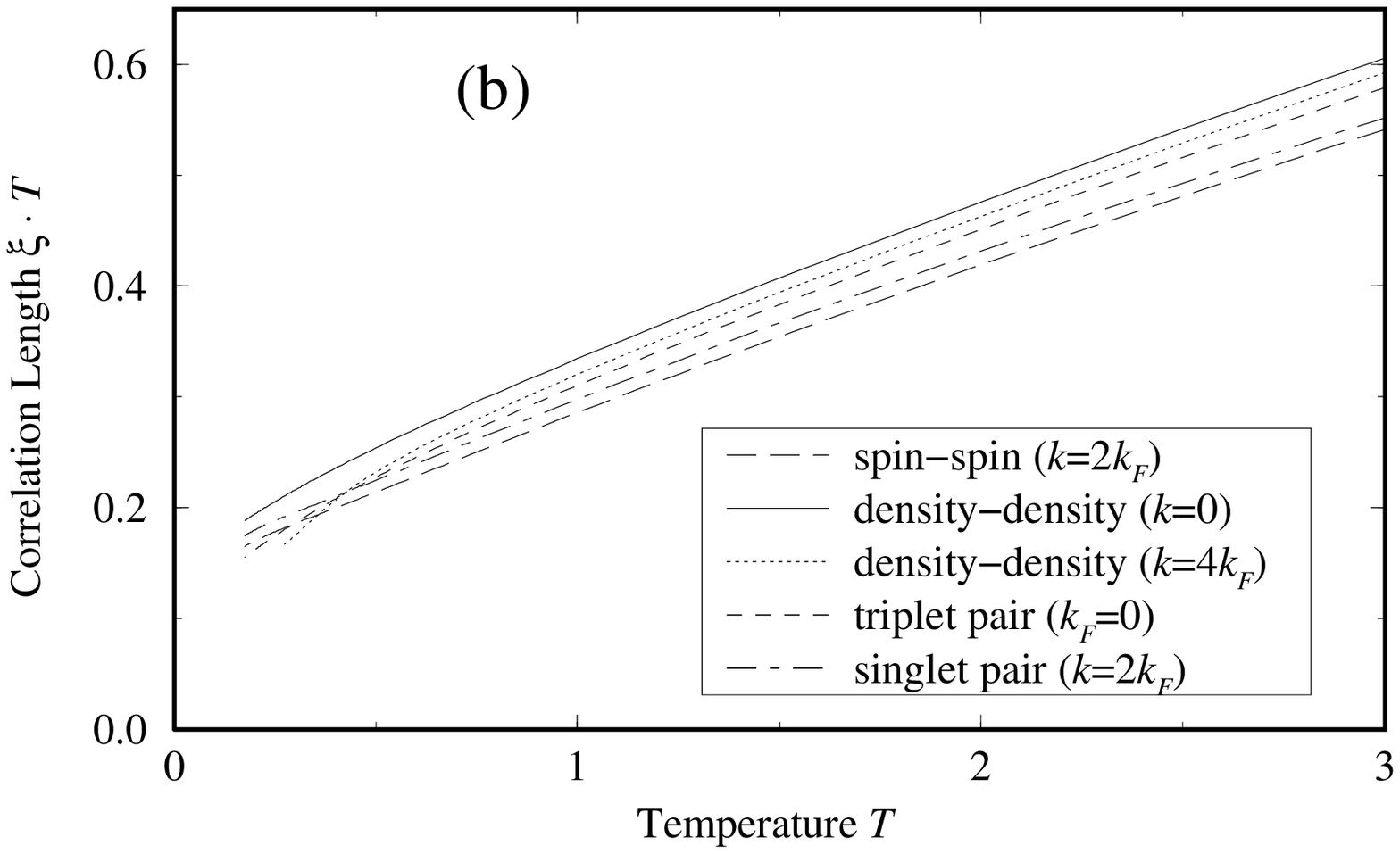}
  \includegraphics[width=0.95\columnwidth]{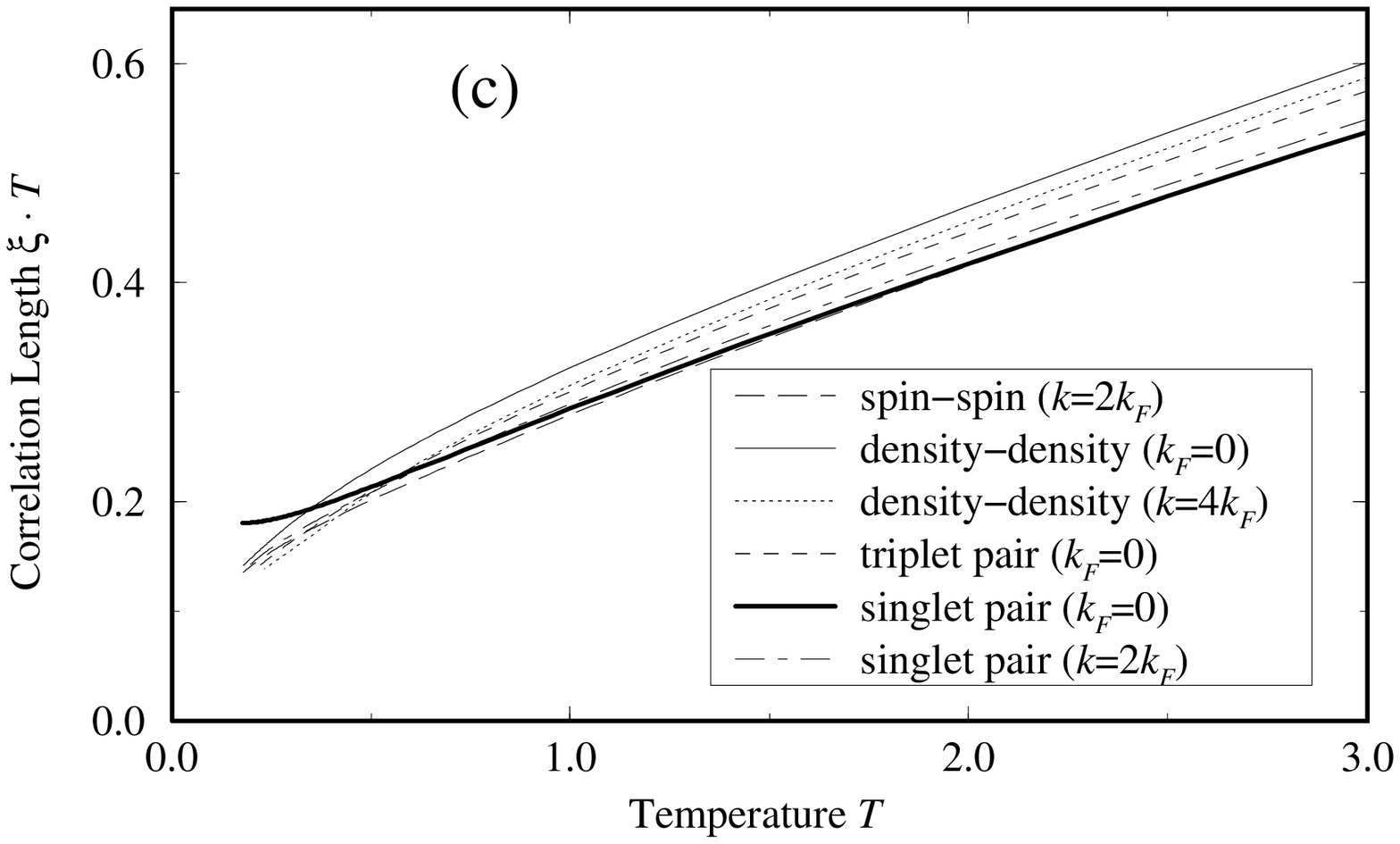} %
  \caption{\label{fig:corr1} Plot of the leading d-d, s-s, sp and
  tp correlation lengths $\xi$ for $X=0.5$ and 
  (a) $\mu=-1$, (b) $\mu=0.4$ and (c) $\mu=0.6$. 
  The corresponding densities $n_0=n(T\to 0)$ read 
  (a) $n_0\approx0.38$, (b) $n_0\approx0.88$ and (c) $n_0=1.19$. 
  The diagrams show the respective correlation $\xi \cdot T$ 
  as a function of $T$. 
  As an example, the inset of (a) depicts the wavevectors $k$ of the
  oscillating CLs. The
  the circles ($\circ$) correspond to the $T\to 0$ limit 
  of $k=2k_F$ and $k=4k_F$ with $k_F=n\pi/2$.}
\end{figure}

Fig.~\ref{fig:corr1} shows the (leading) thermal CLs for three different
chemical potentials ($\mu=-1,0.4,0.6$). The choice of $\mu$ 
covers the TLL ($\mu=-1$) and the LEL phase ($\mu=0.6$) as well as a
point close to the phase transition ($\mu=0.4$). In these figures 
we have plotted $\xi \cdot T$ vs.~$T$ which allows to determine 
the low temperature asymptotics. As $\xi=\gamma/T$ for $T\to
0$ (cf.\ eq.~(\ref{eq:xidiverg})), the curves should become linear and
intersect the ordinate at $\gamma$. According to CFT (see 
Sec.~\ref{xi-smallT}), the factor $\gamma$ depends on the
spin and charge velocities and the critical exponent $K_c$. 

For $\mu=-1$ ($n_0\approx0.38$) in Fig.~\ref{fig:corr1} (a) 
the system is in the TLL regime. The non-oscillating d-d CL
dominates for all temperatures. The leading s-s CL for $T\to 0$ 
shows incommensurable oscillations ($k=2k_F$), whereas the $k=0$ part is
strongly suppressed and not shown in the figure. Additionally, the
$k=4k_F$ d-d correlation lengths dominate the $k=2k_F$ d-d and all s-s
CLs. This scenario can be easily understood by
the fact, that the spin velocity is very small ($v_s\ll v_c$), which can
be verified by Fig.~\ref{fig:sus_05}. The ratio $v_c/v_s\gg 1$ enters
eq.~(\ref{eq:dens_corr_2}), thus these correlations are
suppressed for low temperatures. 
For the same reason, the tp correlation length 
is crossed over by the leading incommensurable sp correlation length 
at low temperatures, cf.\ eq.~(\ref{eq:sp_corr_0}). The inset of
Fig.~\ref{fig:corr1} (a) depicts the wavevectors $k_F$. We have also
plotted the $T\to 0$ values of $2k_F$ and $4k_F$ with $k_F=\frac{\pi}{2} n$
which agree perfectly with the TMRG data.

The case $\mu=0.4$ ($n_0\approx 0.88$) shown in Fig.~\ref{fig:corr1} (b) 
also falls into the TLL regime, but is situated close to the
phase transition to the LEL. The CLs generally get smaller, 
because the charge velocity $v_c$ is decreased 
(Fig.~\ref{fig:sus_05}). Note, that the crossing phenomena of CLs 
become very rich close to the transition. 

A crossover of sp and d-d correlations at finite temperature $T_c\approx
0.5$ is observed in Fig.~\ref{fig:corr1} (c) 
for $\mu=0.6$ ($n_0\approx 1.19$). And in contrast to the latter
cases, the leading sp correlation length is shown to be commensurable.
This agrees with the predictions of eq.~(\ref{eq:LEL_1}) and (\ref{eq:LEL_2}), 
if $K_c>1$ is assumed. 

\begin{figure}
  \centering \includegraphics[width=0.95\columnwidth]{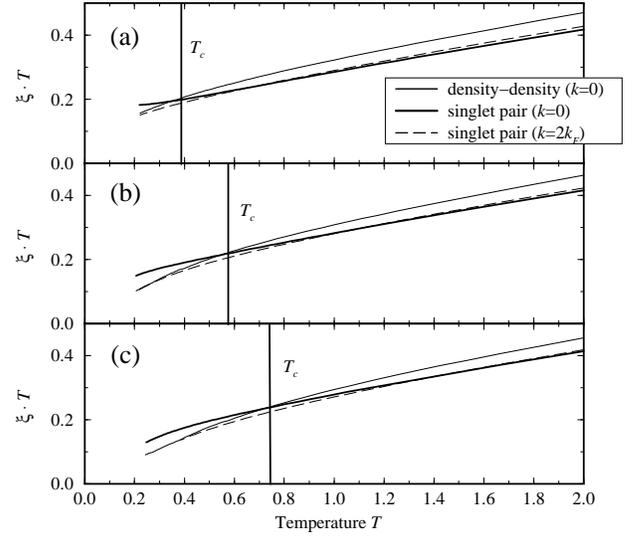}%
  \caption{\label{fig:corr2} Crossing phenomena of density-density
    ($k=0$) and singlet pair ($k=0,2k_F$) correlation lengths for
    various chemical potentials (a) $\mu=0.6$ (b) $\mu=0.8$ and (c)
    $\mu 1.0$. The diagrams plot the particular correlation length 
    $\xi \cdot T$ as a function of temperature $T$. 
    The corresponding densities $n_0=n(T\to 0)$ are given by
    (a) $n_0\approx 1.21$ (b) $n_0\approx 1.92$ (c) $n_0\approx 1.99$. }
\end{figure}

The crossing temperature $T_c$ increases for
higher fillings. This is demonstrated by Fig.~\ref{fig:corr2}, which
shows only the leading d-d and sp correlation lengths for $\mu=0.6,0.8,1.0$.
Note, that the CLs in Fig.~\ref{fig:corr1} (c) and
Fig.~\ref{fig:corr2} do not show asymptotic $\frac{1}{T}$ behaviour
for the achievable temperature region $T>0.2$. 


\section{Conclusions} \label{sec:concl}

In this paper we have studied in detail the thermodynamics of the
bond-charge model by means of numerical TMRG calculations. We focused on
the parameter region $0\le X/t \le 1$, which is representative for the
whole model due to particle-hole symmetry.

Our computations show data for the grand canonical potential, charge
and spin susceptibilities, the specific heat, density and
thermal correlations lengths. The detailed study of small ($X/t=0.1$),
intermediate ($X=0.5$) and large ($X=0.9$) bond-charge coupling was
performed to determine the underlying physics of the model.

In accordance with previous works, 
numerical studies of the spin and charge susceptibilies exhibit two
$T=0$ phases of the model. For less than half-filling we principally find
Tomonaga-Luttinger liquid behaviour, where spin and charge 
excitations are gapless. For more than half-filling a Luther-Emery
liquid phase with spin gap is found, if $X$ is not too large. 

We have shown that the thermodynamics for small $X\ll 1$ essentially 
coincide with that of the Hubbard with an effective Coulomb potential 
$U_\text{eff}$ and hopping amplitude $t_\text{eff}$, as
predicted by bosonisation.  For intermediate coupling
($X=0.5$) the correspondence holds qualitatively, but not
quantitatively. For large $X\approx 1$, the spin gap disappears for all
fillings, which even qualitatively contradicts the bosonisation results.  

Additionally, numerical investigations of the 
specific heat and density 
exhibit interesting phenomena of nearly universal crossing points.  These
can be understood from quite general thermodynamic considerations.

Finally we focused on thermal correlation functions, 
which we characterized by thermal correlation
lengths. In the Luther-Emery liquid phase, we observed interesting
crossing phenomena of superconducting singlet-pair and density-density
correlations at finite temperatures $T_c>0$. 
Even though true superconductivity is not possible in the one-dimensional 
bond-charge model, our studies support a \emph{strong tendency} towards
superconductivity. This indicates, that the bond-charge interaction may
be relevant for superconducting phenomena in higher dimensions. 

As we concentrated on the physics of correlated hopping only,
additional interactions like on-site Coulomb
repulsion $U$ or pair hopping processes $Y$ were not discussed in this paper.
Work is in progress in these matters. 

\begin{acknowledgments}
 This work has been performed within the research program SFB 608 of
 the \emph{Deutsche Forschungsgemeinschaft}. AK is supported by 
 \emph{Studienstiftung des Deutschen Vol\-kes}. We would like to thank
 J.\ Sirker, A.\ Kl\"umper and G.\ Japaridze for valuable discussions.
\end{acknowledgments}

\bibliography{ref}

\end{document}